%% file: LyAlphaNebulae_and_GasAccretion_SV.tex
\documentclass[graybox, envcountchap, authoryear, natbib]{svmult}

\usepackage{mathptmx}       
\usepackage{helvet}         
\usepackage{courier}        
\usepackage{type1cm}        
\usepackage{makeidx}         
\usepackage{graphicx}        
\usepackage{multicol}        
\usepackage[bottom]{footmisc}

\begin{document}

\title*{Gas Accretion and Giant Ly$\alpha$ Nebulae}
\author{Sebastiano Cantalupo}
\institute{Sebastiano Cantalupo \at Institute for Astronomy, ETH Zurich, Wolfgang-Pauli-Strasse 27, CH-8093, Zurich, Switzerland, \email{cantalupo@phys.ethz.ch}}
%
%
\maketitle

\abstract*{
Several decades of observations and discoveries have shown that high-redshift
AGN and massive galaxies are often surrounded by giant Ly$\alpha$ nebulae
extending in some cases up to 500 kpc in size. In this chapter, I review
the properties of the such nebulae discovered at $z>2$ and their connection
with gas flows in and around galaxies and their halos. In particular,
I show how current observations are used to constrain the physical properties
and origin of the emitting gas in terms of the Ly$\alpha$ photon production
processes and kinematical signatures. The emerging picture from these
studies suggest that recombination radiation is the most viable scenario to explain 
the observed Ly$\alpha$ luminosities and Surface Brightness for the large majority
of the nebulae and imply that a significant amount of dense ($n>1$ cm$^{-3}$), ionized 
and cold ($T\sim10^4$ K) "clumps" should be present within and around the halos
of massive galaxies. Spectroscopic studies suggest that, among the giant Ly$\alpha$ nebulae, 
the one associated with radio-loud AGN should have kinematics dominated by 
strong, ionized outflows within at least the inner 30-50\,kpc. Radio-quiet nebulae instead
present more quiescent kinematics compatible with "stationary" situation and, in some
cases, suggestive of rotating structures. However, definitive evidences for accretion
onto galaxies of the gas associated with the giant Ly$\alpha$ emission are not 
unambiguously detected yet. 
Deep surveys currently ongoing using other bright, non-resonant lines such as 
Hydrogen H$\alpha$ and He\,II 1640 will be crucial to search for clearer signatures of
cosmological gas accretion onto galaxies and AGN.
}

\abstract{
Several decades of observations and discoveries have shown that high-redshift
AGN and massive galaxies are often surrounded by giant Ly$\alpha$ nebulae
extending in some cases up to 500 kpc in size. In this chapter, I review
the properties of the such nebulae discovered at $z>2$ and their connection
with gas flows in and around galaxies and their halos. In particular,
I show how current observations are used to constrain the physical properties
and origin of the emitting gas in terms of the Ly$\alpha$ photon production
processes and kinematical signatures. The emerging picture from these
studies suggest that recombination radiation is the most viable scenario to explain 
the observed Ly$\alpha$ luminosities and Surface Brightness for the large majority
of the nebulae and imply that a significant amount of dense ($n>1$ cm$^{-3}$), ionized 
and cold ($T\sim10^4$ K) "clumps" should be present within and around the halos
of massive galaxies. Spectroscopic studies suggest that, among the giant Ly$\alpha$ nebulae, 
the one associated with radio-loud AGN should have kinematics dominated by 
strong, ionized outflows within at least the inner 30-50\,kpc. Radio-quiet nebulae instead
present more quiescent kinematics compatible with "stationary" situation and, in some
cases, suggestive of rotating structures. However, definitive evidences for accretion
onto galaxies of the gas associated with the giant Ly$\alpha$ emission are not 
unambiguously detected yet. 
Deep surveys currently ongoing using other bright, non-resonant lines such as 
Hydrogen H$\alpha$ and He\,II 1640 will be crucial to search for clearer signatures of
cosmological gas accretion onto galaxies and AGN.
}

\section{Introduction}
\label{sec:1}

This chapter reviews the properties of extended Ly$\alpha$ emission in the high redshift universe
(z$>2$) and their connection with gas flows in and around galaxies and their halos. 
In particular, the attention will be focused on giant Ly$\alpha$ nebulae, i.e. emission with
sizes exceeding 100 kpc (unless otherwise noticed, I will always use "physical" units in this chapter)
and in general on Ly$\alpha$ emission extending on scales that are significantly larger than individual
galaxies. In terms of energetics, these systems have integrated Ly$\alpha$ luminosities
($L_{Ly\alpha}$) larger than 10$^{43}$ erg s$^{-1}$. For space reason,
smaller and lower luminosities Ly$\alpha$ halos found around individual 
Lyman $\alpha$ emitters (LAE), e.g. Wisotzki et al. (2016), or in stacking analysis, 
e.g. Steidel et al. (2011) will not be covered in this chapter although 
their origin in some cases may be connected with larger nebulae 
(see e.g., Borisova et al. 2016 for discussion). 
Similarly, I will not review here the detection and study 
of giant nebulae in metal emission line in the low redshift 
universe (e.g., Bergeron et al. 1983, Veilleux et al. 2003). 

The chapter is organized as follows. In section \ref{sec:2}, I provide
an observationally-oriented and historical overview describing
the discovery and characterization of high-redshift Ly$\alpha$
Nebulae. In section \ref{sec:3}, a detailed discussion about the
origin of the Ly$\alpha$ emission is provided in terms of the
atomic processes responsible for the production of Ly$\alpha$
photons. In section \ref{sec:4}, I review the kinematics of the
nebulae and the connection with the physical origin and fate
of the gas including accretion onto galaxies and their halos. 
I summarize the chapter in section \ref{sec:5}. 
Throughout this chapter, a $\Lambda$CDM cosmology with 
$\Omega_m=0.3$, $\Omega_{\Lambda}=0.7$ and H=70 km s$^{-1}$
is assumed.
For reference, one arcsec at z=3 corresponds to about 7.6 kpc with these
cosmological parameters.

\section{Observations of Giant Ly$\alpha$ Nebulae}
\label{sec:2}


\subsection{Quasar Ly$\alpha$ Nebulae}
\label{subsec:2.1}

Being the first and more luminous high-redshift object discovered above z$>$2, quasars have been 
obvious signposts since the mid 1980s to look for putative galactic and gaseous "companions"
in Ly$\alpha$ emission.
One of the first attempt reported in the literature is the pioneering narrow-band observation of 
Djorgovski et al. (1985) on the Lick Observatory 3 meter telescope centred on the radio-loud QSO PKS1614+051
at z$\sim$3.2 that resulted in the discovery of a companion Ly$\alpha$ emitting galaxy (a narrow-line AGN) 
at about 5" from the quasar. 
Later observations by Hu \& Cowie (1987) using the same technique on the 3.6 meter Canada France Hawaii Telescope (CFHT), 
showed the presence of a "bridge" of Ly$\alpha$ emission between the quasar and the companion galaxy. 
In the same year, Schneider et al. (1987) reported the discovery of "companion" Ly$\alpha$ emission to the triply-lensed
radio-loud quasar Q2016+112.

These initial discoveries prompted an intense effort to search for Ly$\alpha$ emission around quasars, mostly of which radio-quiet, 
in subsequent years but despite the large number of quasars observed (about 50), no detectable Ly$\alpha$ candidates were found
(these results were mostly unpublished, see discussion in Hu et al. 1991). The situation changed in the early 1990s when
observational surveys focused on the much smaller sub-sample of radio-loud quasars 
(Hu et al. 1991, Heckman et al. (1991) reported a detection rate of compact or extended companion Ly$\alpha$ emission
close to 100\%. In particular, Heckman et al. (1991) reported the discovery of Ly$\alpha$ Nebulae with sizes of about 100 kpc 
for 15 of the 18 radio-loud quasars observed. The contemporarily discovery of large Ly$\alpha$ Nebulae around 
non-QSO radio-sources (e.g., McCarthy et al. 1987 at z$\sim1.8$) as I will discuss in section \ref{subsec:2.2}, 
suggested to these authors a possible link between the radio and the Ly$\alpha$ emission. At the same time,
searches around radio-loud quasars were also motivated by the possibility to test the hypothesis that
radio-loud quasars and radio galaxies were the same class of objects viewed along different angles with respect
to the radio axis. 

One of the first detection of "companion" Ly$\alpha$ emission to radio-quiet quasars was due to a serendipitous
observation by Steidel, Sargent \& Dickinson (1991) at the Palomar 5 meter telescope: originally searching 
for the continuum counterpart of a z$\sim$0.8 MgII absorber in the spectrum of the radio-quiet QSO Q1548+0917, 
they found instead  a narrow and extended Ly$\alpha$ emission line at the same redshift of the quasar in the the spectrum
of a faint continuum source $\sim$5" away from the QSO. Also, additional narrow-band imaging suggested the
presence of extended Ly$\alpha$ emission around the quasar. In the following year, Bremer et al. (1992) reported
the discovery of extended ($\sim$5") emission in long-slit spectra of two radio-quiet quasars at z$\sim$3.6.
These results showed that "companion" Ly$\alpha$ emission was not restricted to radio-loud quasars only. 

Giant Ly$\alpha$ Nebulae with sizes larger than 10" (or larger than about 100 kpc) 
around radio-quiet quasars remained however elusive for more than two 
decades after the survey of Heckman et al. (1991) around radio-loud quasars.
The only exception was the serendipitous discovery of Bergeron et al. (1999) around the $z\sim2.2$ radio-quiet quasar J2233-606
located in one of the parallel fields of the Hubble Deep Field South. 
The field was observed during science verification of the VLT-UT1 Test Camera using broadband filters and a narrow-band 
filter centred on the quasar Ly$\alpha$ emission. Despite some large-scale residual of the flat-fielding
procedure, the narrow-band image seemed to show extended emission with a maximum projected 
size of about 12" (about 100\,kpc with current cosmological parameters) around the quasar\begin{footnote}{A very recent
and deeper observation with GMOS seems to confirm at least part of this nebula up to a size of about 70\,kpc 
(Arrigoni-Battaia et al, in prep.).}\end{footnote}. On the other hand, during the same years a few individual detections 
of small nebulae extending up to a few arcsec around radio-quiet quasars were reported 
(Fried 1998; M{\o}ller et al. 2000; Bunker et al. 2003; Weidinger et al. 2004) and in some cases associated with
intergalactic gas (e.g., Weidinger et al. 2004, 2005). By the beginning of the 2010a, thanks to long-slit spectroscopic
surveys and small Integral-Field-Unit (IFU) observations, a common picture emerged that associated only relatively small
nebulae (i.e., $<$60-70\,kpc) to about 50\% of radio-quiet quasars between $2<z<5$ 
(Christensen et al. 2006; Courbin et al. 2008; North et al. 2012; Hennawi \& Prochaska 2013; but see Herenz et al. 2015)
However, as I discuss below, these results may have been limited by the small sizes of the IFU Field-of-View (FOV)
and by the use of long-slit spectroscopy that cannot capture the full extent of asymmetric nebulae.

\subsubsection{Rise of the Giants}
\label{subsubsec:2.1.1}

The last few years witnessed a complete revolution in our knowledge of giant Ly$\alpha$ nebulae
around radio-quiet quasars thanks to serendipitous discoveries by means of NB imaging
with custom-built filters (Cantalupo et al. 2014; Martin et al. 2014; Hennawi et al. 2015)
and dedicated surveys with VLT/MUSE (e.g., Borisova et al. 2016). 
Two decades after Hu et al. (1991) and Heckman et al. (1991),
a new narrow-band campaign on quasar fields was initiated in order to search
for "dark galaxies" and fluorescently illuminated intergalactic gas
following the prediction, e.g., of Cantalupo et al. (2005). 
Two pilot programs using a spectroscopic  "multi-slit plus filter" technique (Cantalupo et al. 2007) and deep 
NB imaging ($\sim$20 hours) on FORS/VLT using a custom-built filter for a quasar at z$\sim2.4$
(Cantalupo et al. 2012), revealed a dozen of compact Ly$\alpha$ sources
with no detectable continuum and Equivalent Widths (EW) larger 
than 240$\mathrm{\AA}$, the best candidates for "dark galaxies" illuminated by the quasar.
Circumgalactic gas was also detected in emission extending by several tens of kpc around a few bright galaxies 
but the quasar did not show evidence for extended nebulae, in agreement with previous findings. 

\begin{figure*}[!ht]
\includegraphics[width=\textwidth]{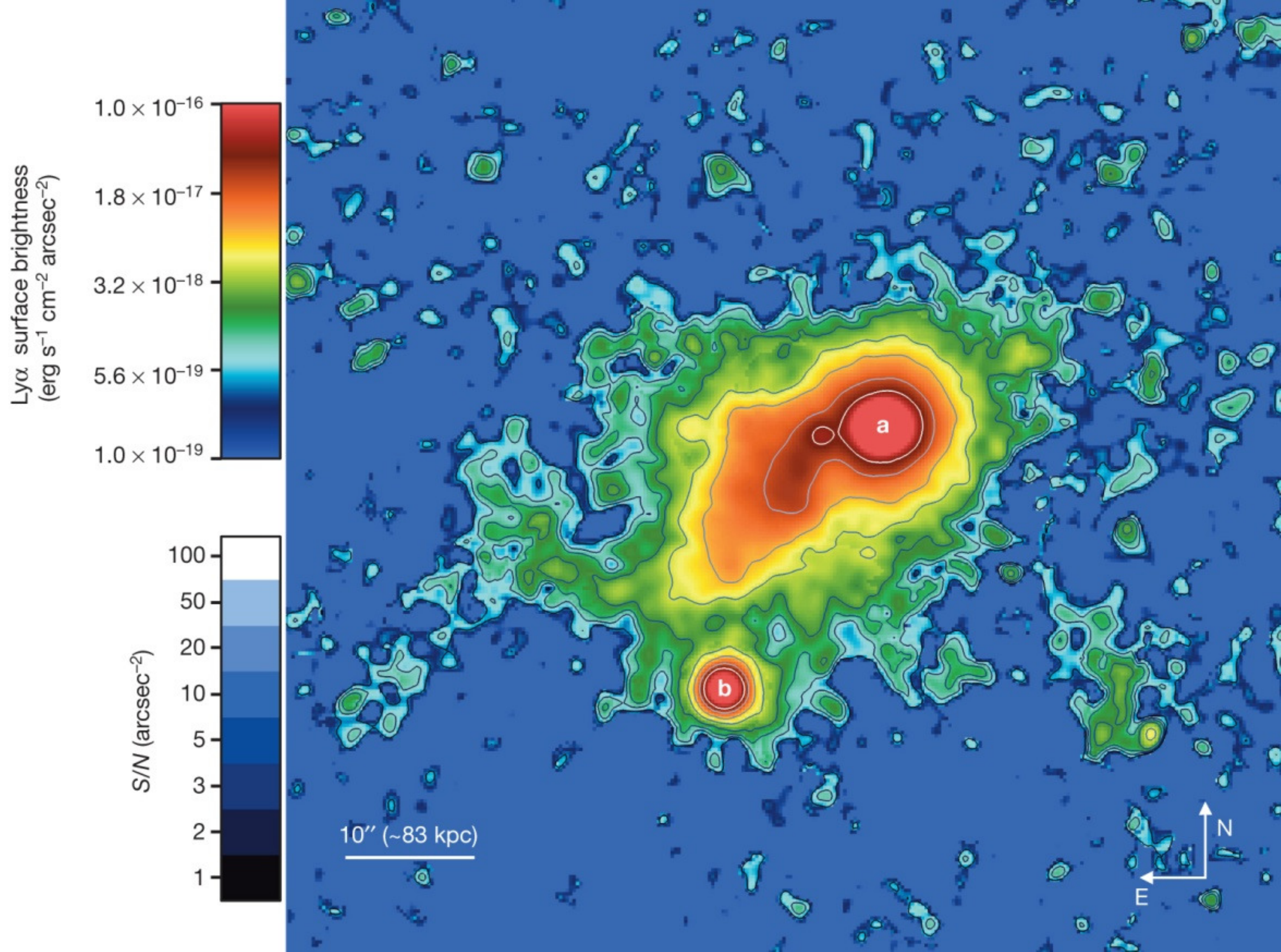}    
\label{FigSlug}
\caption{Continuum-subtracted narrow-band image of the Slug Nebula (Cantalupo et al. 2014) discovered
around the bright, radio-quiet quasar UM287 (labeled ``a'' in the image) in a 10-hour deep 
observation made with a custom filter installed on Keck/LRIS. The nebula shows filamentary emission
extending on a total projected length of about 55 arcsec ($\sim460$ kpc) and is currently 
one of the largest and more luminous Ly$\alpha$ nebulae discovered to date. Its surprising properties
in terms of extension and high values of Surface Brightness are discussed in details in section \ref{subsubsec:2.1.1}.
(Figure reproduced with permission from Cantalupo et al. 2014).}
\end{figure*}

Stimulated by these initial results, Cantalupo et al. (2014) initiated a campaign using Keck/LRIS
and custom-built filters to search for "dark galaxies" around about ten radio-quiet quasars
at $z\sim2$. Surprisingly, the first quasar observed at Keck/LRIS, i.e. UM287, showed 
clear evidences of extended emission over scales larger than 10" after the first 20 minutes
exposure was obtained. At the end of the total integration of 10 hours and detailed data
reduction, a giant Ly$\alpha$ nebula extending to about 55" ($\sim460$ kpc) was
found around this quasar (Cantalupo et al. 2014), see Fig. 1, and named "Slug Nebula" 
(given its morphology and in honor of the mascot of the University of 
California, Santa Cruz). This discovery was surprising for several reasons:
i) despite its association with a radio-quiet quasar, it is at least twice as large as any previously 
detected Ly$\alpha$ Nebulae including the much more common radio-galaxy halos 
and Ly$\alpha$ blobs as discussed below; ii) given its size, it extend well beyond the viral
halo of the quasar into the intergalactic medium; iii) it shows a very high Surface Brightness
over very large scales that cannot be easily explained unless a large gas clumping factor
within intergalactic gas is invoked (see section \ref{subsec:3.1}). During the same night, a second quasar field
was observed using a different custom-built filter, selected among the quasar-pair sample 
of Hennawi \& Prochaska (2013) and showing some hints of extended emission, and another 
giant Ly$\alpha$ nebula with a size of about 300 kpc was discovered (Hennawi et al. 2015). 
The particularity of this discovery included also the presence of a physically associated 
quasar quartet and a large overdensity of Ly$\alpha$ emitting galaxies (differently than the
Slug Nebula). Named "Jackpot Nebula" given the rarity of such systems, it traces likely a
very peculiar region of the Universe and possibly a proto-cluster. In the same year of these
discoveries, Martin et al. (2014) presented the detection of another giant nebula originally
found around one of the six quasars observed in narrow-band imaging as a part of the
Keck Baryonic Structure Survey (e.g., Trainor \& Steidel 2012). Subsequent observations of other quasars
as a part of the same survey and including results obtained on GMOS/Gemini 
(Arrigoni-Battaia et al. 2016) showed once again however that such detection of giant Ly$\alpha$
nebulae are apparently rare, i.e. with a frequency less than 10\%.

The installation of the MUSE Integral-Field-Spectrograph on the Very Large Telescope in 2014 
provided new opportunities for the detection and study of Giant Ly$\alpha$ nebulae
around quasars thanks to its a large FOV of 1'$\times$1' (about $450\times450$ kpc$^2$;
individual spatial elements have a size of 0.2"$\times$0.2")
and because, by design, it does not suffer from either narrow-band filter losses or 
spectroscopic slit losses. Also, the large number of spatial and spectral elements 
allows for better quasar PSF estimation and removal with respect to NB surveys. 
Because accurate systemic redshifts are not needed for 
MUSE observations (as for any other spectroscopic survey) 
any quasar with Ly$\alpha$ redshifted between the blue and red
edges of the MUSE wavelength range ($2.9<$z$<6.5$) can be
observed.
In one of the first exploratory study as a part of the MUSE
Guaranteed Time Observations (GTO), Borisova et al. (2016) 
observed with short total exposure times (1 hour) 
17 of the brightest radio-quiet quasars in the Universe
at $3.1<z<3.7$ and complemented them with two radio-loud quasar
at the same redshifts. 
 The picture emerging from these MUSE observations is very different than that
based on previous surveys, in that giant nebulae with sizes
larger than 100 pkpc are found around essentially every quasar
 above a surface brightness level of about 10$^{-18}$ erg s$^{-1}$ cm$^{-2}$
 arcsec$^{-2}$. 
 The nebulae detected with MUSE present a large range in sizes
and morphologies, ranging from circular nebulae with a projected
diameter of about 110 pkpc to filamentary structures
with a projected linear size of 320 pkpc (see Fig.2). 
 Despite these differences, the circularly averaged SB profiles 
 show a strong similarity between all the giant Quasar Nebulae
 (including the Slug Nebula at z$\sim2$ once corrected for 
 redshift-dimming) with very few 
 exceptions, both in terms of slope and normalization, suggesting a similar
origin for these systems.

\begin{figure*}[!ht]
\includegraphics[width=4.3in]{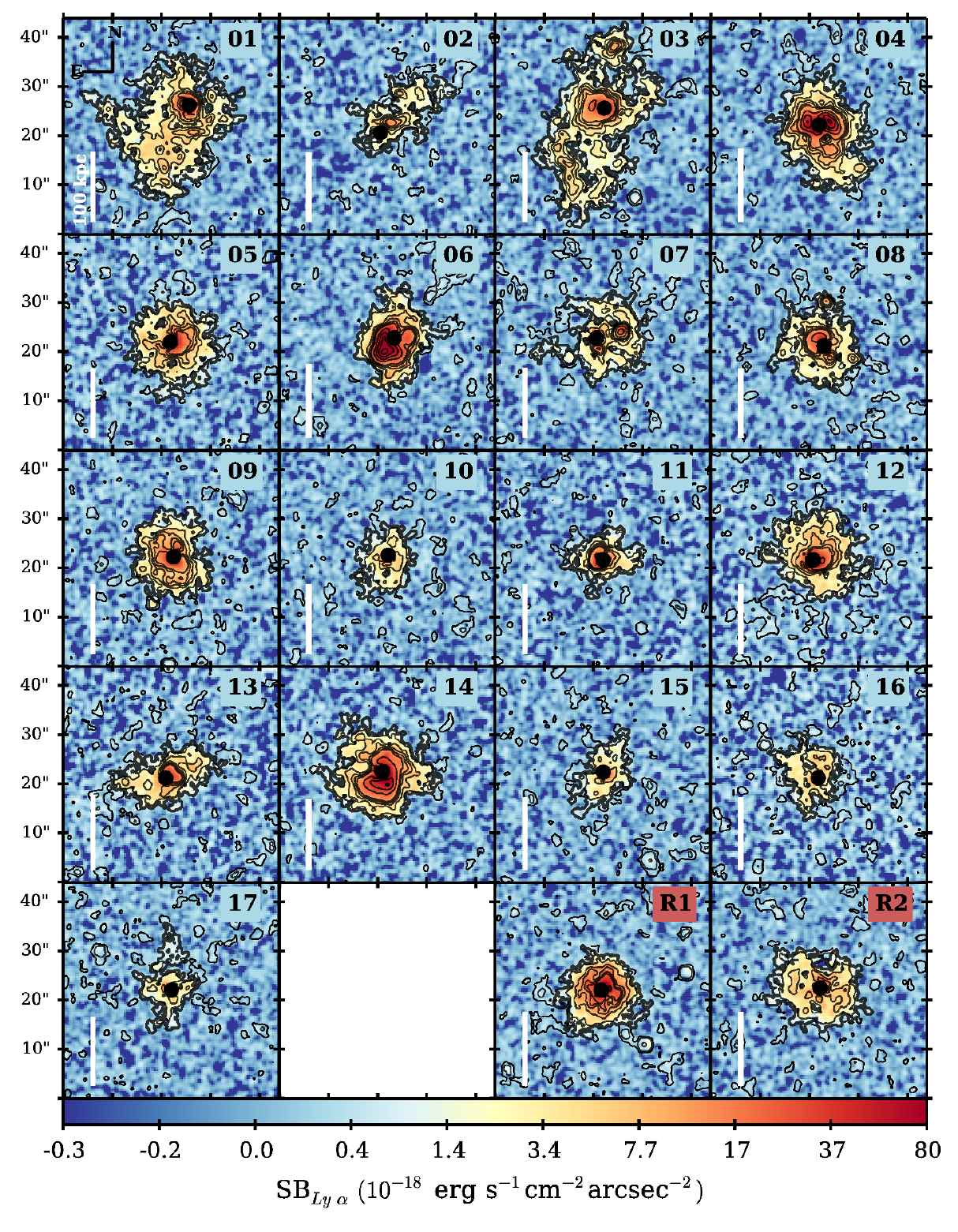} 
\label{FigMQN_SB}
\caption{"Optimally extracted" narrow-band images of the sample of luminous quasars at $3<z<4$ 
observed with MUSE by Borisova et al. (2016) as a part of a "snapshot" survey using short integration
times, i.e. 1 hour per field (see Borisova et al. 2016 for details on the detection and extraction of these
images from the datacubes). The quasars are all radio-quiet with the exception of the two fields labeled 
"R1" and "R2". All nebulae are larger than 100 kpc and extending in some cases up to at least 320 kpc
(e.g., the nebula number 3 or "MQN03") with various morphologies including filamentary structure.
This survey showed that giant Ly$\alpha$ nebulae are ubiquitous around bright quasars, including
radio-quiet ones, in contrast to previous observations at $z\sim2$ as discussed in detail in section \ref{subsec:2.1}.
(Figure reproduced with permission from Borisova et al. 2016).}
\end{figure*}
 
  This 100\% detection rate of giant nebulae around radio-quiet
  quasars obtained with MUSE (see also, Fumagalli et al. 2016) is in stark
  contrast with previous results in the literature as I have reviewed in this section.
  While the asymmetric morphology of the MUSE nebulae may explain 
  the discrepancy with spectroscopic surveys using a single slit position, 
  the difference with the detection rate of NB 
 surveys at z$\sim$2 cannot be completely and easily explained 
  by observational limitations such as NB filter losses, uncertainties in quasar
  systemic redshifts and Quasar Point-Spread-Function (PSF) removal errors.
  Redshift and quasar luminosities may therefore play a role in the appearance
  and properties of Ly$\alpha$ nebulae around quasars and future IFU studies
  extending to lower redshifts (e.g., with the Keck Cosmic Web Imager) and to
  lower quasar luminosities are necessary to properly address these open questions.

\subsection{Radio-galaxy Ly$\alpha$ Halos}
\label{subsec:2.2}

As mentioned in the previous section, follow-up observations of radio-loud sources provided
already in the second half of the 1980s the first evidences for giant Ly$\alpha$ nebulae. 
In this section, I will focus on non-QSO radio sources at z$>1$, i.e. High-z Radio Galaxies (HzRG). 

Although many different classification exists (see e.g. McCarthy 1993 and Antonucci 2012 for a review), the most obvious
distinction between the QSO and non-QSO class of radio source traces its roots from the appearance
of the optical morphology, i.e. from the presence of a quasi-stellar point source versus a
spatially resolved galaxy. Because this distinction may be difficult to be applied at high redshift a more
physical definition can be made instead from the properties of the emission-line spectra. 
Indeed, classical radio-galaxies have spectra with relatively narrow (FWHM$<$2000 km/s) permitted
lines with respect to the broader lines showed by quasars (FWHM$>$5000 km/s). These are sometimes
called "Narrow Line Radio Galaxies" or "type 2". Moreover, high-redshift 
quasars have a much brighter, "thermal" continuum than "type 2" radio-galaxies with a clear non-stellar origin.
It is important to notice however that a small fraction of "spatially resolved" radio galaxies 
show broad lines as well as bright thermal continuum. These are sometimes called "Broad Line Radio Galaxies".
To avoid confusion, in this section and in the reminder of this chapter we will always refer to radio-galaxies
as "type 2".
Like their lower redshift counterparts, HzRG often show extended radio lobes that have been associated
with bi-polar jets (e.g., McCarthy 1987) and show a high degree of polarization. The observation of
relativistic beaming and smaller, one-sided radio-jets in radio-loud quasars have already suggested in the
1980s that radio-galaxies and radio-loud quasars may be part of the same population of AGN but seen
at different orientations (e.g., Barthel 1989; see Antonucci 1993 for a review and, Antonucci 2012 for more
recent discussion). 

Narrow-band imaging and spectroscopy of HzRGs, soon after their discovery in the 1980s and until very recently, 
has produced a large literature of detections and studies of large Ly$\alpha$ nebulae, similarly to radio-loud quasars.
The first observations by McCarthy et al. (1987) of the radio-galaxy 3C 326.1 at z$\sim$1.8 at the Lick Observatory
revealed Ly$\alpha$ emission surrounding the radio lobes and extending by about 70 kpc 
(with current cosmological parameters). Few years later, McCarthy et al. (1990) reported the discovery of the first
giant Ly$\alpha$ nebula extending over 120 kpc around the z$\sim$1.8 radio galaxy 3C 294. This nebula is highly
elongated and well aligned with the inner radio source axis. Extended C\,IV and He\,II emission was also detected.
One of the first detection of extended Ly$\alpha$ emission around a radio-galaxy above redshift of three was
reported by Eales et al. (1993) studying the z$=$3.22 radio galaxy 6C 1232+39, a system with a "classical"
double radio structure oriented along the same direction of the Ly$\alpha$ emission but showing some
differences in the morphology and symmetry of the optical emission with respect to the radio.

Larger sample of HzRG observations were started to be obtained in the 1990s as a part of several campaigns that
resulted in the discovery of several tens of HzRG at z$>$2 (e.g., R\"ottgering et al. 1994). 
The most notably discoveries included a giant Ly$\alpha$ halo extending over 140 kpc around a HzRG at z$\sim3.6$
(van Ojik et al. 1995) showing a complex kinematical structure spatially correlated with the radio jet:
large velocity widths (FWHM$\sim$15000 km/s) in proximity of the radio jets and a aligned, 
low SB component with narrow velocities (FWHM$\sim$250 km/s) extending 40 kpc beyond both sides of the radio source. 
 I will discuss and compare in detail in section \ref{sec:4}, the kinematical properties of radio-loud and radio-quiet nebulae, including this system.

Subsequent observational campaigns found almost ubiquitous detection of extended Ly$\alpha$ emission around HzRG
extending in several cases above 100 kpc in size (e.g., Pentericci et al. 1997, Kurk et al. 2000, Reuland et al. 2003, Villar-Martin et al. 2003,
Venemans et al. 2007, Villar-Martin et al. 2007, Humphrey et al. 2008, Sanchez \& Humphrey 2009, Roche et al. 2014) 
and showing a wealth of morphological structures, including filaments, clumpy regions and cone-shaped
structures. The common features of all these detections include: i) apparent alignment between Ly$\alpha$ emission and 
the radio jet axis (e.g. Villar-Martin et al. 2007), ii) broad kinematics (with some exceptions at the far edge of the nebulae), iii) associated extended C\,IV and
He\,II emission, iv) invariable association with large overdensity as traced by companion Ly$\alpha$ galaxies (e.g., Venemans et al. 2007)
or multi-wavelength observations from X-ray to the infrared (Pentericci et al. 1997). 
In terms of luminosities and surface brightness the HzRG halos show similar values with respect to radio-loud and radio-quiet quasar
nebulae, while UV line ratios and kinematics seems to be quite distinct form the radio-quiet quasar Nebulae (see e.g., Borisova et al. 2016)
as I will discuss in section \ref{sec:4}.

\subsection{Ly$\alpha$ blobs}
\label{subsec:2.3}

The historically-distinct category of giant Ly$\alpha$ Nebulae called "Ly$\alpha$ blobs" made their
appearance with the deep narrow-band observations of Steidel et al. (2000) of a region 
containing a high overdensity of galaxies both in projected and velocity space in the so called 
"Small Selected Area 22$^h$" (SSA22) field (Lilly et al. 1991; Steidel et al. 1998). While searching for candidates Ly$\alpha$ emitting galaxies
in this field at z$=$3.09, Steidel et al. (2000) found to their surprise "two very luminous, very extended regions of line
emission which we descriptively call 'blobs'". Extending by about 15 to 17 arcsec (115 to 132 kpc with current cosmology),
they were among the largest Ly$\alpha$ Nebulae found at that epoch, with similar extent and luminosities to radio-loud quasar and
radio-galaxy Nebulae. However, these "blobs" were apparently lacking any associated, bright continuum or radio source
and therefore considered as a possibly different category of objects. In particular, Steidel et al. (2000) considered the
possibility of a "cooling flow" origin for the emission, by analogy with H$\alpha$ emission from observed cooling-flow
clusters or photo-ionization by heavily obscured, highly star forming galaxies. However, no firm conclusion was possible
with the available data for that time. 
The "mysterious" nature of these Ly$\alpha$ blobs attracted the attention of several theoretical and numerical studies,
right after their discoveries until very recently, that tried to explain the peculiarity of these systems with a variety of physical
explanations that did not require a bright photo-ionizing source such as an AGN. These models ranged from galactic superwinds 
(including e.g., Taniguchi \& Shioya 2000) to cooling radiation from the so called "cold-mode accretion" 
(including, e.g. Haiman et al. 2001; Fardal et al. 2001; Dijkstra \& Loeb 2009) as I discuss more in detail in section \ref{sec:3}.

From an observational and historical point of view, a Ly$\alpha$ "blob" (LAB) could then be defined as an extended Ly$\alpha$ emission
over scales significantly larger than a single galaxy that does not seem to contain an AGN (at the time of their discovery). 
Extended Ly$\alpha$ nebulae around galaxies that did not contain optically bright or radio-loud AGN in overdense fields 
were already known before the observations of Steidel et al. (2000), e.g. Francis et al. (1996) and later re-observations published in 
Francis et al. (2001), and Keel et al. (1999), although in most cases hints of obscured AGN were present in these studies. 
Steidel "blobs" however were the first one in this category to exceed the "giant Ly$\alpha$ Nebula" size of 100 kpc. 
Later Subaru observations of the same SSA22 field made by Matsuda et al. (2004) exploring a much larger area than the
original survey by Steidel et al. (2000) found several extended Ly$\alpha$ nebulae around galaxies, i.e. 33 nebulae with area
exceeding 15 arcsec$^2$ in addition to the Steidel "blobs" (also called LAB1 and LAB2). However, none of those new detections
were close to LAB1 and LAB2 in terms of overall sizes. For instance, the largest new detection by Matsuda et al. (2004) had an
area that was only half the one of LAB2 and a maximum projected size of about 75 kpc. Nevertheless, this overdense field
showed clearly an excess of extended sources with respect to "blank" observation suggesting a possible relation between
galaxy (or AGN) overdensities and extended Ly$\alpha$ emission. Similarly, deep narrow band imaging 
obtained by Palunas et al. (2004) of the overdense field at z$\sim2.38$ found by Francis et al. (2001), 
revealed a large Ly$\alpha$ blob with sizes compatible with Steidel's LAB1 and several smaller nebulae around galaxies.

LAB1 was the first LAB to be studied in detail in multi wavelength observations and revealed the presence of a
strong submillimeter source with a bolometric luminosity in excess of 10$^{13}$L$_{\odot}$ (e.g., 
Chapman et al. 2001, Geach et al. 2005) but no evidence from deep Chandra X-ray observations 
of a clear X-ray counterpart (Chapman et al. 2004).
However, these limits did not exclude the possibility of a luminous AGN but heavily obscured along our line of sight.
LAB2 instead showed, in addition to a (fainter) submillimeter source (Chapman et al. 2001), clear evidences
for hard X-ray emission (e.g, Basu-Zych \& Scharf 2004) and therefore for the presence of a partially 
unobscured AGN. 
This link between LABs and luminous star forming galaxies stimulated new observational campaigns to detect 
LABs by searching, e.g. around luminous infrared sources. During one of such campaigns, Dey et al. (2005)
discovered a 160 kpc Ly$\alpha$ emitting nebula around a luminous mid-infrared source first detected
with the Spitzer Space Telescope. This nebula shared many similarities with the previously detected
LAB1 and LAB2. Although X-ray information was not available at that time, the presence of 
narrow and centrally concentrated C\,IV and He\,II emission within the Nebula suggested again an association
with a Type-2 AGN. Differently from radio-galaxy halos, this nebula showed however relatively narrow
Ly$\alpha$ emission in velocity space and possibly a ordered velocity shear. As we will discuss in section \ref{sec:4},
this is compatible with other nebulae associated with radio-quiet AGN. 

Giant LABs with sizes above 100\,kpc are mainly discovered by targeting known overdense regions or bright
infrared and submillimeter galaxies. Blank-field surveys using narrow-band imaging confirmed that 
giant LABs were indeed extremely rare (e.g. Saito et al. 2006, Yang et al. 2009), i.e. presenting  
a comoving number density less than 10$^{-6}$ Mpc$^3$ (e.g., Yang et al. 2009). Broadband surveys plus
spectroscopic follow-ups were slightly more successfully, detecting one LAB with a size of about 100\,kpc
and three smaller nebulae in a volume of about 10$^{8}$ Mpc$^{3}$ (Prescott, Dey \& Jannuzi 2013).
It is instructive to compare these numbers with the comoving number densities at $2<z<3$ of
bright X-ray selected AGN, i.e. $n_{X}\sim10^{-6}$ Mpc$^{-3}$ for $L_X>10^{44.5}$ erg s$^{-1}$ (Ueda et al. 2003),
optically-bright quasars at z$\sim3$, i.e. $n_{\mathrm QSO}\sim10^{-7}$ Mpc$^{-3}$ for $M_i(z=2)<-26.7$ (Shen et al. 2007),
and HzRG, i.e. $4\times10^{-8}$ Mpc$^{-3}$ for $L_{2.7 GHz} > 10^{33}$ erg s$^{-1}$ Hz$^{-1}$ sr$^{-1}$
(see Venemans et al. 2007). 

Detection of smaller nebulae with sizes up to 40-70\,kpc are less rare in blind narrow-band surveys
and about ten of such discoveries were reported in the past, including Nilsson et al. (2006), Yang et al. (2009), 
Smith et al. (2009), and Prescott et al. (2009). Deeper and larger narrow-band observations around 
the SSA22 fields and including some blank fields over a total volume of about 10$^{6}$ Mpc$^{3}$
as a part of the "Subaru Ly$\alpha$ blob survey" resulted in the discovery of about seven new giant 
LAB with sizes around 100 kpc (Matsuda et al. 2006). An estimation of the number densities of these systems
in this survey however, could be affected by the presence of the very overdense region in SSA22.
In almost all cases, evidences for associated submillimeter or AGN sources were
found at the time of the discovery or with subsequent multi-wavelength observations 
(e.g., Geach et al. 2009, Overzier et al. 2013, Hine et al. 2016; see also Scarlata et al. 2009 and Ao et al. 2015).

Despite the different techniques and volume probed, clustering analysis suggests that 
the sizes of detected LABs could be positively correlated with the environment overdensity,
although the statistic is small (e.g., Yang et al. 2009, Matsuda et al. 2006). 
The very recent detection reported by Cai et al. (2016) of a LAB with a projected size of about 
$\sim$440 kpc at the center of one of the largest overdensity known at $z\sim2.3$ seem to provide
additional support for this suggestion. However, also in this case, there are evidences  
for at least one associated AGN (see also Valentino et al. 2016 for another example).

\ \\

As I have review in this section, several decades of observations and discoveries have produced 
a large literature of extended and giant Ly$\alpha$ nebulae that have been classified in various ways and
with different nomenclatures depending on the technique and target of the original surveys, i.e. quasars, 
radio-galaxies, overdense regions or "apparently blank" fields.  Their comparable volume densities 
and the almost invariable association with AGN or massively star forming galaxies seem to suggest however
that most of these distinction may be artificial. 
There are however some indication that the kinematical 
properties and line ratios (e.g., considering He\,II, C\,IV and Ly$\alpha$) are distinct among 
radio-quiet and radio-loud nebulae, as well as for some LAB, as I will discuss in section \ref{sec:4}.

\section{Origin of the emission}
\label{sec:3}

In this section, I will review the three physical processes that are able to produce extended and bright Ly$\alpha$ emission:
i) recombination radiation following hydrogen photoionization (by a quasar or star forming galaxy), ii) "continuum-pumping"
or Ly$\alpha$ scattering of the photons produced in the quasar broad-line-region or within the Interstellar Medium (ISM)
of a star forming galaxy, and iii) Ly$\alpha$ collisional excitation and recombination radiation following collisional ionization
(so called Ly$\alpha$ cooling radiation). All these mechanisms require the presence of "cool" gas (i.e., with temperatures well below 10$^5$ K). 
In addition, recombination radiation and "continuum-pumping" also require that the gas is "illuminated" by the quasar 
or by some other bright source of UV photons. In this case, the resulting Ly$\alpha$ emission is sometimes called "fluorescent" in the literature
(e.g., Cantalupo et al. 2005, Kollmeier et al. 2010, Cantalupo et al. 2012, 2014; Borisova et al 2016).

\subsection{Recombination radiation}
\label{subsec:3.1}

In this case, Ly$\alpha$ photons are produced in a (photo-)ionized medium as a consequence of radiative recombination cascades
and the Ly$\alpha$ emissivity will be simply proportional to the hydrogen recombination rate times the ionized hydrogen density squared.
Because the hydrogen recombination rate has a relatively mild dependence from temperature, i.e. about linear (e.g., Osterbrock 1989) around 10$^4$ K, 
this is the easiest case to model. The actual number of Ly$\alpha$ photons produced by each recombination event (and therefore by each
ionizing photon, in photoionization equilibrium) will depend on the details of the recombination cascades that populate the 2 P2 state
that will decay to the 1 S2 producing a Ly$\alpha$ photon. A good approximation, even in the low-density Intergalactic Medium,
is to assume that every Lyman-line photon from $n>2$ levels is converted into lower series photons plus either 
Ly$\alpha$ or two-photon continuum (Case B approximation, Baker \& Menzel 1938). In this case, the fraction of 
Ly$\alpha$ photons for each recombination event ranges between 0.68 and 0.61 for $10^4<(T/K)< 10^{4.7}$
(Cantalupo et al. 2008) showing again only a mild dependence on temperature. 
For reference, in the opposite case (Case A) when all Lyman-series photons leave the cloud, the fraction of Ly$\alpha$ photons
for each recombination will be reduced by about a half.
Combining these information with the Case B
hydrogen recombination coefficient, one can obtain the effective Ly$\alpha$ recombination coefficient $\alpha^{eff}_{Ly\alpha}$
for Case B,
that can be approximated with the expression (Cantalupo et al. 2008): 
\begin{equation}
\alpha^{eff}_{Ly\alpha}\sim2\times10^{-13} T_4^{-1.26}\ \mathrm{ cm^3 s^{-1}},
\end{equation}
where, $T_4=10^4K$, for $10^3<(T/K)< 10^{4}$. For a fully ionized cloud of hydrogen and helium, the Ly$\alpha$ integrated volume emissivity is therefore:
\begin{equation}
4\pi j_{Ly\alpha}\sim3.7\times10^{-24} n^2 T_4^{-1.26}\ \mathrm{ erg\ s^{-1} cm^{-3},}
\end{equation}
 where $n$ is the gas density in units of atoms per cm$^{-3}$.
For a uniform gas slab with depth $L$ in units of kpc at redshift z, the observed surface emissivity - equivalent to the surface brightness ignoring radiative transfer effects - will be given therefore by:
\begin{equation}
SB^{em}_{Ly\alpha}\sim8\times 10^{-17} n^2 T_4^{-1.26} L \cdot [(1+z)/4]^{-4}\ \mathrm{erg\ s^{-1} cm^{-2} arcsec^{-2}}.
\end{equation}
The observed values of the SB of giant Ly$\alpha$ nebulae range from a few times $10^{-18}$
 erg s$^{-1}$ cm$^{-2}$ arcsec$^{-2}$ in the external parts (at distances of 50-100\,kpc from the central sources or the central peak of the Ly$\alpha$ emission)
 to  $10^{-16}$ erg s$^{-1}$ cm$^{-2}$ arcsec$^{-2}$ in the central regions.
From this simple calculation, it is easy to understand that the observed SB would inevitably require large gas densities ($n>0.1$) in order to be explained by
 pure recombination emission. These high density estimates were obtained and claimed since the discoveries of the first nebulae around radio-loud quasars
 and galaxies (e.g. Heckman et al. 1991). 
 
 Are these densities compatible with our picture of cold gas within massive halos? It is instructive to rewrite the previous expression 
 rewriting the density $n$ in terms of the baryonic overdensity with respect to the canonical value of 200 for
 collapsed objects ($\delta_{200}$). 
 Making use of the definition of clumping factor, i.e. $C_{L}=<n^2>_L/<n>_L^2$ where the average is made over a cylinder with depth L
 and projected area of 1 arcsec$^2$ on the sky, we obtain:
 \begin{equation}
 SB^{em}_{Ly\alpha}\sim5\times10^{-20} C_L <\delta_{200}>^2_{L_{100}}  L_{100} T_4^{-1.26}\cdot [(1+z)/4]^2\ \mathrm{ erg\ s^{-1} cm^{-2} arcsec^{-2}},
 \end{equation}
 where $L_{100}=L/100$\,kpc. In this formula, no assumptions have been made about the density distribution and density inhomogeneities are encoded
 in the clumping factor term $C_L$. Assuming that 100\,kpc is the typical projected length of the cold gas distribution in the halo,
 it is then straightforward to see that, even if all the halo gas is in the form of a fully ionized and uniform medium at $T=10^4$K, 
 the predicted SB will be smaller than the observed values by several orders of magnitude. 
 In this case, the observed SB constrains the clumping factor on scales corresponding to 1 arcsec$^2$ on the sky plane, i.e. on a square with
 side of about 7 kpc at $z\sim3$ and on 100\,kpc along the line of sight to be at least $C_L\sim20-2000$. This simple estimates is confirmed
 by detailed radiative transfer and cosmological simulations, e.g. Cantalupo et al. (2014).
 For a series of small clouds with uniform density in a empty medium, the clumping factor is equal to the inverse of the volume filling factor $f_v$, 
 and therefore the implied values of $f_v$ are smaller than $10^{-3}$ and likely much lower (i.e., $10^{-4}-10^{-5}$), considering that, 
 if a hot medium is present as it should be in these massive halos, most of the mass and density will be in the hot component rather 
 than the cold one used for the estimate above. These small filling factors would imply again that the individual densities of the clouds 
 should be higher than $n > 0.1$ and likely closer to $n\sim 1-10$ cm$^{-3}$ considering the presence of a hot medium. 
 
  A similar analysis can be made considering the total Ly$\alpha$ luminosity, see e.g. the classical approach 
  used by McCarthy et al. (1990). In this case, integrating the Ly$\alpha$ emissivity given above one would simply obtain: 
\begin{equation}
  L_{Ly\alpha}\sim 3.7\times10^{-24} n^2 T_4^{-1.26} f_v V\ \mathrm{ erg\ s^{-1}},
  \end{equation}
  where V is the total volume in cm$^3$, or more conveniently expressed in terms of the halo virial masses:
  \begin{equation}
  L_{Ly\alpha}\sim 2.6\times10^{47} n^2 T_4^{-1.26} f_v M_{12} \cdot [(1+z)/4]^{-3}\ \ \mathrm{ erg\ s^{-1}},
  \end{equation}
  where $M_{12}$ is the virial mass in units of $10^{12} M_{\odot}$.
  In this case, the value of $f_v$ in the literature is typically fixed to the one derived from
  extended line emission in local AGN (10$^{-6}$ to 10$^{-4}$, e.g. Heckman et al. 1989). McCarthy et al. (1990)
  adopted $f_v$ to 10$^{-5}$ "as a guess", as reported in their paper, and this value was then adopted later in the literature
  as a sort of "standard" value. As discussed above, our SB consideration lead to similar values. Once $f_v$ is fixed, 
  the individual densities of the clouds are therefore constrained to be of the order of 1-100 cm$^3$ given the Ly$\alpha$
  luminosities obtained by the various studies in the literature ($10^{43}<L_{Ly\alpha}/$(erg s$^{-1}$)$<10^{45}$)
  and assuming that the nebulae are associated with massive halos ($1<M_{12}<10$).
  Interestingly enough, similar densities are required in the pure recombination case to explain the reported 
  limits on He\,II emission compared to Ly$\alpha$ for some cases, e.g. for the Slug Nebula (Arrigoni-Battaia et al. 2015) 
  and for the MUSE radio-quiet quasar nebulae (Borisova et al. 2016). 
  
  The analysis above shows that recombination radiation is a viable scenario to explain the observed Ly$\alpha$ luminosities
  and SB if the nebulae are composed by dense ($n>1$ cm$^{-3}$), highly ionized and cold ($T\sim10^4$) clumps with volume filling
  factors smaller than $10^{-3}$ or, analogously, if the gas clumping factor is larger than about a thousand on projected scales with size
  of about 5-8 kpc. 
   Assuming that such a population of clouds - not resolved by current cosmological simulation - do exists within massive halos,
   we should examine if they can be indeed ionized by the observed sources within or around the Ly$\alpha$ Nebulae.  
   The required ionizing luminosities Q$_{ion}$ are simply derived using photoionization equilibrium
   and considering that about 68\% of recombinations produce Ly$\alpha$ photons (Case B) at $T\sim10^4$ as discussed above:
   \begin{equation}
   Q_{ion}\sim f_c^{-1} L_{Ly\alpha}\times10^{11}\ \mathrm{ photons\ s^{-1}},
   \end{equation}
    where $f_c$ is the covering factor of the gas as seen by the 
   ionizing source. Assuming $f_c\sim1$ this would imply that the required ionization luminosities to explain the Ly$\alpha$ emission 
   of the giant Ly$\alpha$ nebulae ($10^{43}<L_{Ly\alpha}/$(erg s$^{-1}$)$<10^{45}$) are in the range $10^{54}-10^{56}$ s$^{-1}$.
   These ionization luminosity could be explained easily by AGN, even with modest luminosities, or by starbursts with star formation
   rates of at least a few hundred to few thousand solar masses per year (depending on the ionizing photon escape fraction from the galaxy ISM
   and on the stellar Initial Mass Function, e.g. Leitherer et al. 1999). As a reference, the quasar associated with the Slug Nebula, the largest
   and more luminous Ly$\alpha$ nebula discovered to date (Cantalupo et al. 2014) has an expected $Q_{ion}\sim10^{57.5}$ photons s$^{-1}$
   using the measured rest-frame UV luminosity and assuming a standard for this type of quasars (e.g., Lusso et al. 2015).  
   Given the ubiquity of AGN and massively star forming galaxies (traced, e.g. in submm) within or around giant Ly$\alpha$ nebulae, there are 
   therefore enough ionizing photons for the recombination scenario. In section \ref{sec:4}, I will review the possible origin of the illuminated gas
   in terms of kinematical signatures of inflows or outflows.

\subsection{Continuum pumping (scattering)}
\label{subsec:3.2}

 Due to the resonant nature of the Ly$\alpha$ emission, the detected photons from extended nebulae may also "originate" from the ISM
 of embedded galaxies or from the broad line region of the associated AGN, rather than being produced "in situ" by recombination
 processes. A "scattering" scenario for some nebulae and in particular for LABs have been proposed in the past as a possible
 origin of the level of polarization detected in some sources, e.g. in SSA22-LAB1 by Hayes et al. (2011) and Beck et al. (2016), and in
 a radio-galaxy nebula by Humphrey et al. (2013). In other cases, radio-quiet nebulae imaged in polarization failed to show
 any detectable level of polarization, e.g. Prescott et a. (2011). Theoretical prediction for the level of polarization produced by
 scattering by a central sources have been presented, e.g. by Dijkstra \& Loeb (2008) using idealized geometries and
 velocity fields and extended recently to radiative hydrodynamics simulations by Trebitsch et al. (2016). These models
 suggest that scattering from a central source is able to produce a larger and steeper polarization profile with respect
 to scattering by photons produced by "in situ" processes, such as collisional excitation. Comparison with the data
 cannot clearly exclude one of these two scenarios but current data, at least for SSA22-LAB1, 
 seems more compatible with a "in situ" Ly$\alpha$ photon production rather than scattering from a "central" source.
 
 In case a significant fraction of Ly$\alpha$ photons detected in the extended nebulae are produced by scattering 
 from the "central" source, then all the estimates of masses and densities based on the assumption of pure recombination
 emission discussed above should be revised. The "scattering" contribution will be of course related to the density
 of neutral hydrogen atoms ($n_{H^0}$) rather than the ionized density squared as in the recombination case. 
 The Ly$\alpha$ optical depth at line center is given by:
 \begin{equation}
 \tau_0\sim3.3\times10^{-14} T_4^{-1/2} N(HI) ,
 \end{equation}
 where $N(HI)$ is the neutral hydrogen column density in units of cm$^{-2}$.
 As an example, using the estimated hydrogen ionization rates of the Slug Nebula quasar (UM287), 
 i.e. $\Gamma\sim7.7\times10^{-9}$ at distances of 100 kpc, we can derive $N(HI)$ in the simple case of uniform 
 small clouds with individual densities of $n$ and filling factor $f_v$ exposed to ionizing radiation 
 as discussed in the previous section. In photoionization equilibrium, the averaged neutral fraction of hydrogen
 along a path of length L from the quasar will be given by:
 \begin{equation}
 <x^{eq}>_L=L^{-1} \int_0^L x^{eq} dL' \sim L^{-1} \int_0^L n\alpha(T) \Gamma^{-1} dL' \sim 10^{-9} L^2 n,
 \end{equation}
 where the last term is valid at $T\sim10^4$K,
 $\alpha(T)$ is the hydrogen recombination coefficient ($\sim3\times10^{-13}$ at $T_4\sim1$),
 L is the length in units of kpc.
 Because the $N(HI)$ column density up to a length $L$ is given in this case 
 by $N(HI,L)=<x^{eq}>_L\cdot n \cdot f_v \cdot L$, then the Ly$\alpha$ optical depth at line center 
 is simply:
 \begin{equation}
 \tau_0\sim10^{-6} (L/1kpc)^3 (n/1cm^{-2})^2 (f_v/10^{-5}).   
 \end{equation}
 Assuming that most of the scatterings are produced at line center,
 then a continuum photon produced by the AGN is only scattered into Ly$\alpha$ at distances
 larger than 100\,kpc from the source, even for large clump densities. 
 These distances are only reduced by about a factor of two for sources that are ten times fainter
 than UM287. Because most of the emission is concentrated in the inner parts, scattering contribution
 in the case the gas is illuminated by the ionizing radiation of a source should be therefore 
 negligible and recombination radiation should be the dominant mechanism. 
 
 The only case in which scattering may be the dominant production mechanism of the large Ly$\alpha$
 emission of the giant nebulae is when the central source is not highly ionizing the gas but still producing 
 copious amounts of Ly$\alpha$ and continuum photons slightly blueward of Ly$\alpha$. In this case, 
 studied numerically by, e.g. Cantalupo et al. (2014), the gas will be mostly neutral and the optical 
 depths will become very large. 
 The scattering will be extremely efficient and
 the main problem with this scenario would be to transport out to hundred kpc scales, as observed for the
 Ly$\alpha$ Nebula, the photons that are resonantly trapped in the inner regions. During scattering, the
 Ly$\alpha$ photons will perform a random walk both in space and in frequency. Unless the parameters
 are fine tuned, most of the scatterings in space will be very close to each other and the photons will
 escape, with a single fly-out, only when scattered in the wings of the line spectral distribution. 
 In order to transport out these photons to another scattering location and increase the size of the Nebula, 
 the column densities need to be large enough to produce scattering in the wings of the line, 
 i.e. $N(HI)>10^{21} cm^{-2}$, resulting in much larger column densities and masses of neutral gas
 with respect to what observed in absorption around quasars (e.g. Prochaska et al. 2013; see
 also Cantalupo et al. 2014 for discussion). 
 Although scattering and a mostly "neutral" scenario for the gas seems less plausible than recombination
 emission to explain the origin of giant Ly$\alpha$ nebulae, 
 observations of non-resonant lines such He\,II 1640 and hydrogen H$\alpha$ from giant Ly$\alpha$
 Nebulae may be used in the next future to better constrain the scattering contribution to the emission.

\subsection{Collisional Excitation (cooling)}
\label{subsec:3.3}

When both neutral hydrogen and free electrons are present and the electron temperature is around
a few times $10^4K$, the collisional excitation rate of Ly$\alpha$ ($q_{Ly\alpha}$) 
may be several orders of magnitude larger than the Ly$\alpha$ effective recombination rate (see, e.g. Cantalupo et al. 2008)
and therefore produce in principle very strong Ly$\alpha$ emission from lower density gas. 
Differently than the recombination case, however, the collisional excitation rates are a very strong 
function of temperature, with an exponential decline for temperatures lower than a few times $10^4$K
dropping strongly below the recombination rate just below $T\sim10^4$K. On the other hand,
when the electron temperature approaches $10^5$K collisional ionization dominates over collisional
excitations. As a result, collisional excitation is a very efficient process to produce Ly$\alpha$ photons
in a partially ionized medium only for electron temperatures around $2-5\times10^4$K. 
In absence of photoionization and heating sources, e.g. in collisional ionization equilibrium, radiative
losses (cooling) due to collisional excited Ly$\alpha$ would quickly reduce the electron temperature
effectively suppressing further Ly$\alpha$ emission. A source of heating is therefore required
in order to balance the Ly$\alpha$ cooling effects. On the other hand, if heating brings electron temperatures close to 
$10^5$K, Ly$\alpha$ collisional excitation would be suppressed again, therefore the heating source
should be fine tuned in order to produce a stable supply of Ly$\alpha$ emission with collisional excitation. 

Since the discovery of the Steidel's LABs in SSA22, a numerous series of theoretical and numerical papers
have addressed the possibility that these nebulae - apparently lacking a source of ionization - could be
powered instead by Ly$\alpha$ cooling (including, e.g., Haiman et al 2001, Fardal et al. 2001, Furlanetto et al. 2005,
Dijkstra \& Loeb 2009, Goerdt et al. 2010, Faucher-Giguere et al. 2010, Rosdahl \& Blaizot 2012). 
The suggested source of heat in all these studies is given by the conversion of the gravitational potential energy
due to cosmological accretion of gas into massive dark matter halos. Alternative sources of energy such as
galactic superwinds have been also proposed by Taniguchi \& Shioya (2000) with a similar energy budget
with respect to the gravitational accretion case but also sharing the same limitations as discussed below.
 
A simple estimate of the energy associated with cosmological gas accretion for a Navarro-Frenk-White profile 
with concentration parameter $c=5$ gives (Faucher-Giguere et al. 2010):
\begin{equation}
<\dot{E}_{grav}>\sim3.8\times10^{43} M_{12}^{1.8} [(1+z)/4]^{3.5}\ \mathrm{erg\ s^{-1}}.
\end{equation}
For Ly$\alpha$ nebulae associated with very massive halos with $M\sim10^{13} M_{\odot}$, this order of magnitude 
estimate for the potential energy is interestingly close to the observed Ly$\alpha$ emission. In our standard picture
for the formation of these massive halos, we think however that a significant fraction of this potential energy
should result in the formation of a hot gaseous halo with temperatures of the order of $T\sim10^7$K as a result
of the virial shocks. At these temperatures, the gas would more slowly lose energy by other radiation channels 
rather than Ly$\alpha$, e.g., free-free continuum radiation in the Extreme UV and X-ray, and, once the temperature
decreases, further energy may be radiated away via line emission from He\,II or from heavier elements, if present.
Analytical and numerical models in the past ten years have nonetheless suggested that some fraction of the densest
accreting material could not form a stable shock and therefore could cool and produce Ly$\alpha$ 
(e.g. Dekel \& Birnboim 2006). Cooling Ly$\alpha$ emission models have therefore introduced an efficiency factor
that takes into account how much of the potential energy is dissipated by this cold gas phase 
and required that this efficiency factor is at least few tens of percent in order to obtain enough 
Ly$\alpha$ photons. Unfortunately, it is very difficult to predict the value of this efficiency factor as it would
depend on the details of accretion and cosmological structure formation and on a very accurate estimate
of the ionization and temperature state of the gas.  
 Depending on these details, many questions remain open about how and where within the halo the potential
 energy (or the energy produced by galactic superwinds) could be dissipate in form of Ly$\alpha$. 
 
 The majority of the models discussed so far ignore however that sources of ionization do exist within all
 nebulae observed to date. In most cases, as I have reviewed in section \ref{sec:2}, these sources are AGN or massively
 star forming galaxies. The estimated ionization rates show that there are enough ionizing photons to power the
 nebulae and therefore it seems current clearer that a pure Ly$\alpha$ cooling origin of the nebulae 
 is not necessarily required by the majority of the observations. 
 This may be confirmed by duty cycle arguments and comparing the number densities of
 quasars, radio-galaxies and overdense regions where LABs are found as discussed in previous section 
 (see also Borisova et al. 2016 and Overzier et al 2013 for further discussion). Questions however remain
 about the possible contribution of collisional Ly$\alpha$ emission to recombination radiation as this would
 have important consequences on the estimates of densities and masses of the nebulae. 
 Addressing these questions would likely require a new generation of theoretical and numerical models able
 to resolve the physics and the small scales associated with dense regions within massive halos and,
 at the same time, the interplay between the gas and the local radiation sources 
 (e.g., in terms of photoionization, photo-heating and feedback).

\section{Origin of the emitting gas, kinematics and gas flows}
\label{sec:4}

 In the previous sections, we have seen that extended Ly$\alpha$ emission 
 is a common phenomenon around bright high redshift quasars (with possible differences below $z\sim3$ for 
 radio-quiet systems), radio-galaxies and in overdense regions of the Universe. The ubiquity of 
 AGN and massively star forming galaxies associated with these nebulae and simple analytical considerations discussed in 
 section \ref{sec:3} suggest that most of the emission may be due to recombination radiation from 
 dense and cold clouds within the nebulae. Because the emission is sensitive to gas density squared,
 the emitting gas could be a small fraction of the total gas in and around the massive halos associated
 with these systems, both in terms of volume and mass. In this section, we will look at the kinematical signatures
 derived from Ly$\alpha$ spectra of these sources in order to address the questions about the
 possible origin and fate of this cold gas component. 
 
 The proposed origins for the emitting gas and its relation to gas flows in the literature include: i) cosmological accretion 
 from the IGM, ii) outflowing material from galaxy and AGN feedback, iii) in-situ formation from hot gas condensation.
 The suggested options about the fate of the gas include: i) accretion or "recycling" into galaxies (inflow) 
 or into gaseous disks (rotation), ii) expulsion from the galaxy halos towards the IGM, 
 iii) disruption and thermalization into the hot halo gas.  
 Each of these hypothesis could potentially leave an imprint into the observed gas kinematics in terms of Ly$\alpha$
 spectral profile shapes, velocity shears and velocity dispersion.  
 However, it is important to stress that Ly$\alpha$ line is a resonant line and therefore any spectral information 
 may also reflect complex radiative transfer effects rather than kinematics, especially if the gas is not highly
 ionized as it may be the case for a Ly$\alpha$ cooling origin of the emission.
 
 The first spectroscopic measurements on giant Ly$\alpha$ nebulae were obtained 
 on radio-galaxies and radio-loud quasar nebulae by McCarthy et al. (1987), McCarthy et al. (1990), 
 Heckman et al. (1991b), McCarthy et al. (1996). The properties of these systems from a kinematical point of view appeared remarkably similar:
 the Full Width Half Maximum (FWHM) of the Ly$\alpha$ emission presented large values typically around 1000-1500 km s$^{-1}$
 for all radio-loud nebulae. No hints of velocity shear in excess of 500 km s$^{-1}$ were found, although radio-galaxy nebulae
 seemed to show a more complex kinematical structure. Given the alignment effect between the radio-jets and Ly$\alpha$ emission
 discussed in section \ref{subsec:2.2}, these large FWHM were interpreted preferentially as being associated with powerful outflows from 
 these AGN as a result of the jet-gas interaction (e.g., van Ojik et al. 1997, Humprey et al. 2006).
  However, it was early recognized that such large velocity widths could have been
 also caused by gravitational motion in very massive halos (e.g., Heckman et al. 1991b).
  More recent spectroscopic observations of radio-loud systems confirmed these large FWHM (e.g., Villar-Martin et al. 2003, Humphrey
 et al. 2006, Villar-Martin et al. 2007, Roche et al. 2014)
 but found that some radio-galaxies show, in addition, extended and lower-SB halos that appear more kinematically quiet,
 e.g. with FWHM$\sim$500 km s$^{-1}$. Moreover, some of these extended "quiescent" regions show indication for velocity shifts 
 of few hundred km s$^{-1}$ that in some cases have been interpreted as a possible sign of rotating disks
 (e.g., Villar-Martin et al. 2007) or cosmological gas infall (e.g., Humphrey et al. 2007).
 These models are based on the correlation seen in several radio-galaxy nebulae
 i.e. that the more redshifted side of the nebula is the brighter in both Ly$\alpha$ and radio flux,
 and on the interpretation that this side of the radio-lobes and associated nebula is closer to the observer.
 This is based on the fact that the radio jet directed towards the observer will be Doppler-boosted
 and on the assumption that the Ly$\alpha$ emitting gas closer to the observer will be less 
 "absorbed", i.e. scattered to a larger projected area, by neutral gas in the halo.   
 The relative redshift of this near-side relative to the observer would then imply that the
 emitting nebula has a significant component of infall towards the galaxy. 
 
 Kinematical signatures in radio-galaxy and radio-loud quasars are however difficult to interpret because
 of the complex interaction between the possibly accreting gas and the massive energy input
 of the radio jet. The analysis of extended, rest-frame optical emission lines from several  
 HzRG, e.g. Nesvadba et al. (2006), suggests indeed that kinematics in the brightest parts of the
 radio-loud nebulae should be dominated by powerful outflows. These observations, together
 with the measured line ratios using, He\,II, C\,IV, O\,III and S\,II with respect to H$\alpha$, H$\beta$ and
 Ly$\alpha$, all suggest that this outflowing material is metal rich (solar or super-solar),
 dusty (A(H$\beta)\sim$1-4 mag) and very dense ($n_e\sim500$ cm$^{-3}$) (see e.g., 
 Villar-Martin et al. 2003, Nesvabda et al. 2008). 
 
 Long-slit and integral field spectroscopy of radio-quiet nebulae show
 instead that the large majority of these systems have much smaller FWHM, i.e. FWHM$\sim300-500$ km s$^{-1}$
 than radio-loud systems (e.g., Weidinger et al. 2005, Christensen et al. 2006, Arrigoni-Battaia et al. 2015,
 Borisova et al. 2016), unless they are associated with very overdense environment
 like in the case of SSA22-LAB1 and SSA22-LAB2 (e.g., Matsuda et al. 2006) or in the case
 of the "Jackpot" Nebula (Hennawi et al. 2015). These smaller velocity widths are indicative 
 of better environments where signs of infall or rotation could be studied more in detail without
 the need to disentangle them from the broad component associated with the radio-jets
 and powerful outflows. 
 
 In particular, some radio-quiet giant Ly$\alpha$ Nebulae, such as the Slug Nebula (Cantalupo et al. 2014)
 are extended by several hundred of kpc and therefore the kinematics in their external parts are less likely to be 
 "contaminated" by AGN outflows, if present. Long-slit spectroscopy (Arrigoni-Battaia et al. 2015) and
 integral field observations (Martin et al. 2015) have revealed relatively narrow Ly$\alpha$ emission (FWHM$<350km$ s$^{-1}$) 
 at distances larger than 100\,kpc from the quasar and apparent velocity shears of about 800 km s$^{-1}$
 that seem coherent over the brightest part of the Nebula (i.e., over 200 kpc in projected length). 
 These velocity shears have been interpreted as evidence for rotation and would imply therefore 
 that a massive, gaseous disk-like structure with size of about 150\,kpc could be present in the brightest part of the Nebula
 (Martin et al. 2015) while the more tenuous, extended part have been interpreted as signatures of gas 
 infall from the Intergalactic Medium. In particular, the velocity profile seems consistent 
 with the rotation curve of a disk within a NFW profile of a massive halo but only if the
 "center" of the disk-like structure is located 25" away from the main quasar, in a region of low SB.
 Considering only the high SB regions, the spectra could be also interpreted however as
 arising from projection effects of two different structures located at about 500-800 km s$^{-1}$
 in velocity space away from each other. The presence of a relatively bright continuum source
 in the "redshifted" part of the velocity profile could support this interpretation. Ongoing observations of other
 emission lines such as He\,II, C\,IV (e.g., with MUSE) and H$\alpha$ (e.g., with Keck/MOSFIRE) 
 will help disentangling these two possibilities. 

 Other spectroscopic observations of radio-quiet nebulae (e.g., Martin et al. 2014, Prescott et al.  
 2015) have found signatures of coherent velocity fields over scales of several tens of kpc
 that have been interpreted as evidence of infall (e.g., Martin et al. 2014) or large scale
 rotation in a disk (Prescott et al. 2015). In particular, the observations of Prescott et al. (2015)
 of coherent velocity shears of about $\sim$500 km s$^{-1}$ in both Ly$\alpha$ and in 
 non-resonant lines such as He\,II within the central 50\,kpc of a 80\,kpc-sized 
 Ly$\alpha$ nebula at $z\sim1.67$ provide evidences that the extended gas 
 in this system is produced in situ by recombination radiation 
 and settled in a rotating disk that is at least partially stable against collapse. 
 In a larger spectroscopic study of eight small radio-quiet nebulae at z$\sim2.3$ including
 rest-frame optical emission lines (mostly OIII and H$\alpha$),
 Yang et al. (2014) found instead no significant evidences for bulk motions
 such as inflow, rotation or outflows and suggested that the gas should be 
 "stationary" or slowly outflowing at speed less than 250 km s$^{-1}$ with 
 respect to the central galaxies in these systems. 

\begin{figure*}
\includegraphics[width=4.3in]{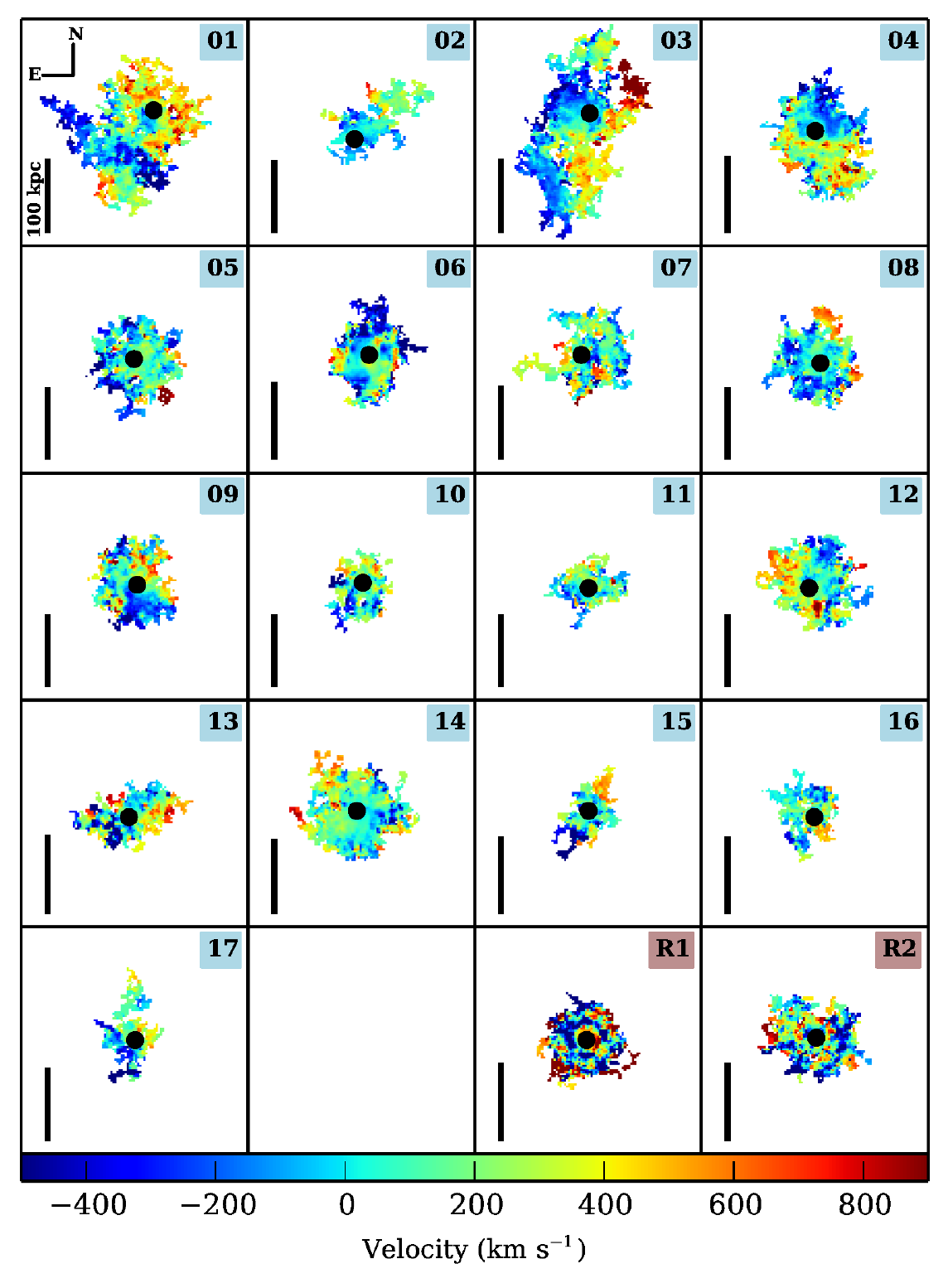}
\label{FigMQN_vel}
\caption{"Velocity maps" of the MUSE Quasar Nebulae presented in Fig. 2 obtained
from the first moment of the flux distribution (see Borisova et al. 2016 for details). This
is the largest sample to date of kinematical maps of giant Ly$\alpha$ nebulae obtained with
integral-field-spectroscopy. As discussed in section \ref{sec:4}, while some systems (e.g. MQN15)
 show possible evidences of rotation in a disk-like structure the majority of the nebulae do not show 
 clear evidences of rotation or other ordered kinematic patterns. Several nebulae 
 show instead coherent kinematical structures over scales as large as 100\,kpc (e.g. MQN01 and MQN03).
 (Figure reproduced with permission from Borisova et al. 2016).}
\end{figure*}

\begin{figure*}
\includegraphics[width=4.3in]{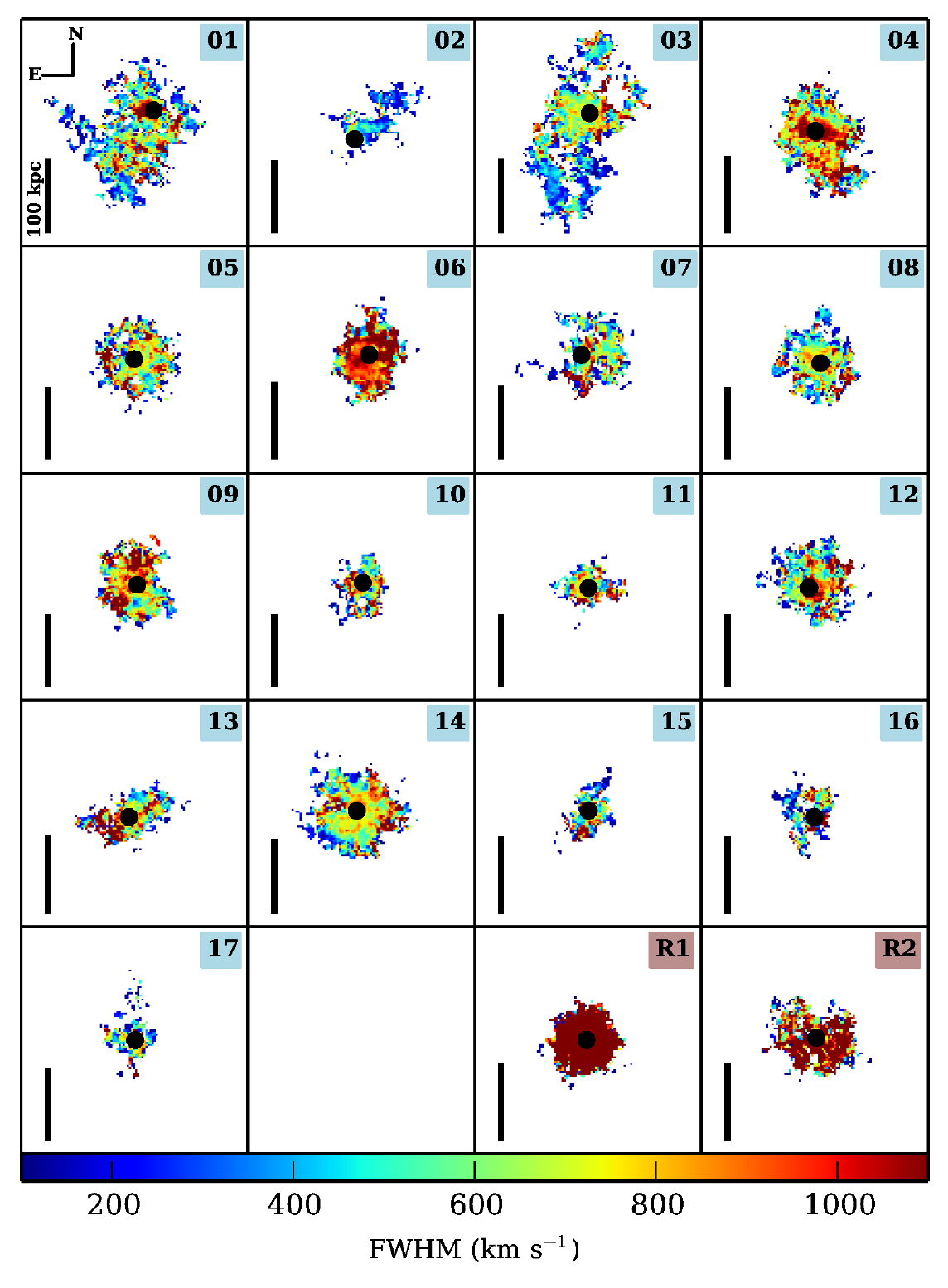}
\label{FigMQN_FWHM}
\caption{"Velocity dispersion maps" of the MUSE Quasar Nebulae presented in Fig. 2 obtained
from the second moment of the flux distribution and expressed in terms of Gaussian-equivalent 
FWHM (see Borisova et al. 2016 for details). This figure shows that the large majority of 
radio-quiet nebulae are narrower in Ly$\alpha$ emission (FWHM$\sim500$ km s$^{-1}$) than radio-loud systems,
i.e. MQN-R1 and MQN-R2 (FWHM$>1000$ km s$^{-1}$), in agreement with the overall kinematical results
discussed in section \ref{sec:4}.
(Figure reproduced with permission from Borisova et al. 2016).}
\end{figure*}

The MUSE observations of about 17 bright radio-quiet quasars and 2 
radio-loud quasars at $3<z<4$ presented by Borisova et al. (2016) 
provided the first large statistical sample of giant ($>100$ kpc) Ly$\alpha$ nebulae 
with full kinematical information from integral field spectroscopy over their 
full detectable extent. Figures 3 and 4
show, respectively, the maps of the first and second moment of the flux distribution, 
i.e. the flux-weighted velocity centroid shift and the dispersion relative to the peak of the integrated 
Ly$\alpha$ emission for each of the MUSE Quasar Nebulae (MQN) (figures taken from Borisova et al. 2016).
While some systems, e.g. MQN15, show possible evidences of rotation in a disk-like structure
with a velocity shear of about 800 km s$^{-1}$, 
the majority of the MQN do not show clear evidences of rotation or other ordered kinematic patterns.
Several MQN, especially the largest ones, show instead coherent kinematical structures 
over scales as large as 100\,kpc, e.g. MQN01 and MQN03. The velocity dispersions,
expressed in terms of Gaussian-equivalent FWHM, clearly shows the main difference 
between radio-quiet and radio-loud systems: the large majority of radio-quiet nebulae are 
narrower in Ly$\alpha$ emission (FWHM$\sim500$ km s$^{-1}$) than radio-loud systems 
(FWHM$>1000$ km s$^{-1}$) in agreement with previous results discussed in this section. 
The only exception is MQN06 but this nebula is peculiar and more similar to radio-loud nebulae 
in terms of all the properties studied in Borisova et al. (2016), including the SB profiles and
higher He\,II/Ly$\alpha$ and C\,IV/Ly$\alpha$ ratios with respect to the non detection for
radio-quiet systems. 

The emerging picture from these observations seems to suggest therefore that
kinematics in radio-loud nebulae may be dominated by ionized outflows of relatively
cold and metal-enriched material within at least the inner 30-50\,kpc from the AGN, 
while, on average, the ionized and clumpy gas in radio-quiet nebulae may be in 
a more "stationary" situation and in some cases settled in a possibly rotating structure. 
Clear evidences from Ly$\alpha$ emission for gas accretion into galaxies from these cold gas 
reservoirs are not currently detected, either because "washed-out" by Ly$\alpha$ radiative transfer effects
or because their magnitude and projection effects could make them to small to be
detected with current facilities. Future deep surveys using other bright, non-resonant lines
such as hydrogen H$\alpha$ or He\,II 1640 would be extremely helpful to search for
small velocity shears and therefore for signature of cosmological gas accretion onto
galaxies and AGN.

\section{Summary}
\label{sec:5}

Several decades of observations and discoveries have produced an extensive
literature of large and giant Ly$\alpha$ nebulae. Depending on the technique
and original target of the observation, i.e. quasars, radio-galaxies, overdense regions
or "blank" fields, they have been classified in various ways and with different
nomenclatures. For historical and practical reasons, I have divided them in quasar Ly$\alpha$ nebulae,
radio-galaxy halos and Ly$\alpha$ Blobs (LAB) but, as discussed in section \ref{sec:2}, 
their comparable volume densities, luminosities and their almost invariable association 
with AGN or massively star forming galaxies suggest however that these nebulae
could be just apparently different manifestations of the same phenomenon. 

Among the three processes associated with the production of Ly$\alpha$
photons (recombination radiation, "scattering" and collisional excitation; see section \ref{sec:3}), 
recombination radiation currently appears as the most viable scenario to explain 
the observed Ly$\alpha$ luminosities and Surface Brightness for the large majority
of the nebulae. If contribution from scattering and Ly$\alpha$ collisional excitation
is negligible, this would imply that the emitting gas should be in the form of
dense ($n>1$ cm$^{-3}$), highly ionized and cold ($T\sim10^4$) structures ("clumps")
 with volume filling  factors smaller than $10^{-3}$ or, analogously, with clumping factors 
 larger than about a thousand on projected scales with size of about 5-8 kpc.
 The apparent ubiquity of giant Ly$\alpha$ nebulae around bright AGN, at least at $3<z<4$,
 and in overdense environment suggests that this cold gas should be a common occurrence
 within and around the halos of massive galaxies. 
 Deep observations of non-resonant lines such He\,II 1640 and hydrogen H$\alpha$ from giant Ly$\alpha$
 nebulae can better constrain the scattering and collisional contribution
 to the emission but their are currently lacking for the majority of the nebulae. Ongoing surveys
 (e.g., with MUSE, MOSFIRE, KMOS and JWST in the future) will soon provide these data
 and therefore potentially help in refining our understanding of the physical properties
 of this cold gas.
 
 Ly$\alpha$ integral-field and long-list spectroscopy shows that radio-loud nebulae
 (i.e. associated with radio-galaxies and radio-loud quasars) are almost invariably associated 
 with larger velocity widths with respect to the majority of radio-quiet systems (see section \ref{sec:4}).
 Together with the analysis of the He\,II/Ly$\alpha$ and C\,IV/Ly$\alpha$ ratios, 
 these observations seem to suggest that kinematics in radio-loud nebulae may 
 be dominated by ionized outflows of relatively cold and metal-enriched material within 
 at least the inner 30-50\,kpc from the AGN. On the other hand, these observations
 suggest that ionized and clumpy gas in radio-quiet nebulae should be in a more "stationary" situation 
 and in only in some cases there are possible evidences that this gas is settled in rotating structures.
 Definitive evidences for the accretion of this cold gas into galaxies from Ly$\alpha$ emission  
 are not clearly detected in the current data. However, it is important to notice that accretion 
 signatures could be "washed-out" by Ly$\alpha$ radiative transfer effects 
 or too small to be detected with current facilities because of their magnitude 
 and possible projection effects. 
 Again, future deep surveys using other bright, non-resonant lines such as hydrogen H$\alpha$ 
 or He\,II 1640 would be extremely helpful to search for small velocity shears and 
 therefore for clearer signature of cosmological gas accretion onto galaxies and AGN.

\begin{acknowledgement}
I gratefully acknowledge support from Swiss National Foundation grant PP00P2\_163824.
\end{acknowledgement}
%

\input{referenc-cantalupo}

\end{document}

%% file: referenc-cantalupo.tex
%
%
%